\documentclass[11pt,a4paper]{article} 
\usepackage{jcappub} %

\usepackage[utf8]{inputenc}
\usepackage{subfig}
\usepackage{amsmath,amssymb}
\usepackage{graphicx}

  \newcommand{\ud}{\mathrm{d}}
  
  \newcommand{\Mpl}{M_{\mathrm{pl}}}
  \newcommand{\lfh}{\lambda_{\phi h}}

  \newcommand{\xif}{\xi}    
    
\def\beq{\begin{equation}}
\def\eeq{\end{equation}}
\def\baq{\begin{eqnarray}}
\def\eaq{\end{eqnarray}}

\title{Destabilization of the EW vacuum in non-minimally coupled inflation}
\author{Stanislav Rusak}
\emailAdd{stanislav.rusak@su.se}
\affiliation{Nordita, KTH Royal Institute of Technology and Stockholm University, Roslagstullsbacken 23, SE-106 91 Stockholm, Sweden}
\abstract{Current cosmological data favour inflationary models non-minimally coupled to gravity. In this work we study the implications of the metastability of the electroweak vacuum in this framework. We consider an inflaton field with a non-minimal coupling to Ricci curvature and a portal coupling between the Higgs field and the inflaton and find that the ratio of the two couplings is severely constrained from stability during inflation and reheating dynamics, with constraints becoming more severe for weaker non-minimal coupling as during inflation Higgs fluctuations are amplified by inflationary expansion and after inflation by parametric resonance due to an oscillating inflaton background.}

\begin{document}

\maketitle

\section{Introduction}

Since its discovery in 2012~\cite{Chatrchyan:2012xdj, Aad:2012tfa} the Higgs field has remained the subject of a vibrant area of research, both in the context of collider physics and wider implications for the universe as a whole. One of the most interesting aspects of Higgs physics is that the values of Standard Model parameters are such that our universe appears\footnote{The stability of the electroweak vacuum depends sensitively on the SM parameters, most notably the mass of the top quark, and it is possible that the current vacuum is indeed absolutely stable up to the Planck scale if the top mass is sufficiently below its best fit value~\cite{Alekhin:2012py,Bednyakov:2015sca}.} to occupy a metastable vacuum state which will eventually decay into a true vacuum at high field values~\cite{Buttazzo:2013uya,Bezrukov:2012sa}. This aspect has received much attention in recent years, especially in the context of early universe cosmology (for a recent review, see~\cite{Markkanen:2018pdo}).

While the lifetime of our metastable vacuum far exceeds the age the universe, it remains unclear how this energetically disfavoured situation was reached. During high-scale inflation, fluctuations in light fields are amplified by \emph{deSitter} dynamics and as such the Higgs field would have been driven to its true vacuum~\cite{Espinosa:2007qp,Lebedev:2012sy,Enqvist:2014bua,Espinosa:2015qea}. This suggests that either the scale of inflation must not have been very high or that some new physics come to the rescue and stabilize the Higgs during inflation. One can stabilize the Higgs potential by introducing new fields which change the running of the Higgs self-coupling at high energies~\cite{EliasMiro:2012ay,Lebedev:2012zw}. A particularly minimal approach is to couple the Higgs fields to the inflaton which is already necessitated by inflation or to couple it non-minimally to gravity~\cite{Herranen:2014cua}. The coupling to the inflaton can stabilize the vacuum by either mixing with  the Higgs at low energies~\cite{Ema:2017ckf} or by inducing a large effective mass for the Higgs at inflationary scales~\cite{Lebedev:2012sy}. In the latter case, while stabilization is achieved during inflation, the vacuum may again become destabilized during reheating as the very same couplings which stabilized the field result in copious production of Higgs quanta due to parametric and tachyonic resonance of the oscillating inflaton field~\citep{Enqvist:2016mqj,Ema:2016kpf, Ema:2017loe, Figueroa:2017slm}.

Many of the earlier works assumed a simple quadratic model for the inflaton which has become increasingly disfavoured by cosmological measurements~\cite{Akrami:2018odb}. Non-minimal couplings are expected on general grounds as they generically appear via radiative corrections even if absent at tree level~\cite{Callan:1970ze,Bunch:1980br,Bunch:1980bs,Birrell:1982ix,Markkanen:2013nwa}. In this work we study the stability of the EW vacuum in the non-minimally coupled quartic model which is in excellent agreement with the data. We focus on the quadratic portal coupling between the inflaton and the Higgs and find that vacuum stability during inflation and reheating impose severe constraints on the couplings. Thus, the stability of the electroweak vacuum may, in fact, prove an important criterion in discriminating between inflationary models.

This paper is organized as follows. In Section~\ref{sec:sec2} we introduce the model and review the inflationary dynamics arising from a scalar non-minimally coupled to gravity. In Section~\ref{sec:during} we study the behaviour of the Higgs during inflation and present constraints arising from stability in the inflationary epoch. In Section~\ref{sec:after} we examine Higgs dynamics during the oscillatory phase of the inflaton and derive constraints arising from parametric resonance. We present the numerical calculations in Section~\ref{sec:numerical} and conclude in Section~\ref{sec:conclusions} with a discussion of the results.

\section{Non-minimally coupled inflation}
\label{sec:sec2}

We consider a model with a quartic inflaton potential where the inflaton is non-minimally coupled to gravity. This setup is similar to the Higgs inflation\footnote{For a recent review of Higgs inflation see~\cite{Rubio:2018ogq}.} scenario~\cite{Bezrukov:2007ep} only now inflation is driven by a new scalar while the Higgs remains a spectator. The action in the Jordan frame is

\begin{eqnarray}
 \label{eq:action1}
 S  =  \int \ud^4 x \sqrt{-g_J}\left[  \frac{\Mpl^2}{2}\left(1+\xif\frac{\phi_J^2}{\Mpl^2}\right)R_J - \frac{1}{2}\partial_\mu\phi_J\partial^\mu\phi_J - \frac{\lambda_\phi}{4}\phi_J^4  -\frac{1}{2}\partial_\mu h\partial^\mu h - V_{h\phi}\right]
\end{eqnarray}
which can be transformed into the Einstein frame by a conformal transformation $g_{\mu\nu} \equiv \Omega^2g_{\mu\nu}^{J}$ with $\Omega^2 \equiv \left(1+\xif\frac{\phi_J^2}{\Mpl^2}\right)$ and the field redefinition

\begin{equation}
  \ud \phi \equiv \sqrt{\frac{\Omega^2 + 6\xif^2\phi_J^2/\Mpl^2}{\Omega^4}}\ud \phi_J
\end{equation}
to make the kinetic term of the inflaton canonical. This can be integrated to give~\cite{GarciaBellido:2008ab}

\begin{equation}
\label{eq:phiJtophiM}
  \sqrt{\xif}\frac{\phi}{\Mpl} = \sqrt{1+6\xif}\operatorname{arsinh}\left(\sqrt{\xif(1+6\xif)}\frac{\phi_J}{\Mpl}\right) - \sqrt{6\xif}\operatorname{arsinh}\left(\sqrt{\frac{6\xif^2}{1+\xif \phi_J^2/\Mpl^2}}\frac{\phi_J}{\Mpl}\right).
\end{equation}
The mapping between frames gives rise to three separate regimes for the inflaton. These can be identified as 

\begin{eqnarray}
 \frac{\phi}{\Mpl} \simeq \left\{\begin{array}{ll}
       \frac{\phi_J}{\Mpl}, & \qquad \phi_J^2 \ll \frac{\Mpl^2}{6\xif^2}
       \\
       \mathrm{sign}(\phi_J)\sqrt{\frac{3}{2}}\xif \left(\frac{\phi_J}{\Mpl}\right)^2, & \qquad \frac{\Mpl^2}{6\xif^2}  \ll \phi_J^2 \ll \frac{\Mpl^2}{\xif}
       \\
     \mathrm{sign}(\phi_J)  \sqrt{\frac{3}{2}}\log\left[1+\xif\left(\frac{\phi_J}{\Mpl}\right)^2\right], & \qquad \phi_J^2 \gg \frac{\Mpl^2}{\xif}
    \end{array}\right. .
\end{eqnarray}
Correspondingly there are three separate regimes for the inflaton potential in the Einstein frame -- quartic, quadratic, and exponentially flat: 

\begin{eqnarray}
\label{eq:potential}
 V_E(\phi) \simeq \left\{\begin{array}{ll}
       \frac{\lambda}{4}\phi^4, & \qquad \phi_J^2 \ll \frac{\Mpl^2}{6\xif^2}
       \\
      \frac{\lambda\Mpl^2}{6\xif^2}\phi^2, & \qquad \frac{\Mpl^2}{6\xif^2}  \ll \phi_J^2 \ll \frac{\Mpl^2}{\xif}
       \\
       \frac{\lambda\Mpl^4}{4\xif^2}\left(1-e^{-\sqrt{\frac{2}{3}}\frac{|\phi|}{\Mpl}}\right)^2, & \qquad \phi_J^2 \gg \frac{\Mpl^2}{\xif}
    \end{array}\right. .
\end{eqnarray}
Inflation happens in the large-field plateau $\sqrt{\xif}\phi_J \gg \Mpl$ and the inflationary observables can be shown to be\footnote{This is in fact equivalent to the Starobinsky model~\cite{Starobinsky:1980te} $\mathcal L = \Mpl^2R/2+\mu R^2$ with $\mu = \frac{\xif^2}{4\lambda_\phi}$.}

\begin{equation}
 \mathcal P_\zeta = \frac{V_*}{24\pi^2\epsilon_*\Mpl^4} \simeq \frac{\lambda N^2}{72\pi^2\xi^2}, \qquad n_s = 1-\frac{2}{N}, \qquad r = \frac{12}{N^2}.
\end{equation}
These values are in excellent agreement with the current cosmological constraints and the tensor-to-scalar ratio lies in the region that will be probed by the next-generation polarization experiments\footnote{However, even those future constraints can be evaded in Palatini gravity~\cite{Rasanen:2017ivk}.}~\cite{Errard:2015cxa}. The measured amplitude of the power spectrum~\cite{Akrami:2018odb} constraints the ratio $\lambda/\xi^2 \approx 5\times 10^{-10}$.

\section{\label{sec:during}Higgs during inflation}

If the Standard Model is valid all the way up to inflationary scales then the Higgs is a light field and therefore its fluctuations obtain a spectrum $\mathcal P_{\delta h}\sim H_\mathrm{inf}^2/(2\pi)^2$ on superhorizon scales. Furthermore, the renormalization group running of the Higgs self coupling predicts that $\lambda_h$ becomes negative around the energy scale $\mu_c \sim 10^{10}$ GeV.  In order for the vacuum to not be destabilized by inflationary amplification of fluctuations, the Higgs must remain heavy during inflation\footnote{Strictly speaking this requirement can be relaxed somewhat -- for a light Higgs the average field value performs a random walk which may end up subcritical in our observable patch~\cite{Espinosa:2015qea}. However, this depends on the initial field value and the duration of inflation. We take the heaviness of the Higgs as a conservative bound.}. This is achieved through coupling to the inflaton which has a large amplitude during inflation. The equation of motion for the Higgs in the Einstein frame is

\begin{equation}
  \ddot h + (3H - \partial_t\log \Omega^2)\dot h - \frac{\nabla^2}{a^2}h + \Omega^{-2}V_h' = 0.
\end{equation}
During slow-roll inflation we can estimate

\begin{equation}
  \left|H^{-1}\partial_t\log\Omega^2\right| = \sqrt{\frac{2}{3}}\frac{\dot\phi}{H\Mpl}\simeq \sqrt{\frac{4\epsilon}{3}}.
\end{equation}
In non-minimally coupled inflation we have $\epsilon\sim \eta^2 \sim 1/N^2$ so that the additional friction term coming from the conformal transformation is unimportant\footnote{Note that this is the case during inflation, but not necessarily during reheating. We shall discuss this issue further in the next section.}.

\subsection{Radiative corrections}

We must take care that the addition of the Higgs does not spoil the successful inflationary predictions. In particular, quantum corrections from Higgs loops must not dominate over the inflationary potential. The effective mass of the canonical variable $\tilde h \equiv h/\Omega$ is 

\begin{equation}
  m_h^2\equiv V_{\tilde h\tilde h} = \lfh \left(\frac{\phi_J}{\Omega}\right)^2 - \frac{1}{4}\left(\partial_t \log\Omega^2\right)^2 + \frac{1}{2}\partial_t^2\log\Omega^2
\end{equation}
and the corresponding Coleman-Weinberg correction due to Higgs loops is
\begin{equation}
   V_\mathrm{CW} = \frac{m_h^4}{64\pi^2}\log\left(\frac{m_h^2}{\mu^2}\right).
\end{equation}
From the slow-roll evolution we can estimate

\begin{equation}
  m_\Omega^2 \equiv \frac{1}{2}\partial_t^2\log\Omega^2 - \frac{1}{4}(\partial_t\log\Omega^2)^2 \simeq - \frac{\lambda_\phi\Mpl^2}{9\xif^2}e^{-2\sqrt{\frac{2}{3}}\frac{\phi}{\Mpl}}
\end{equation}
Therefore the potential induced by the conformal factor alone, $V\sim m_\Omega^4$, is suppressed compared to the inflationary potential by $e^{-2\sqrt{\frac{2}{3}}\phi/\Mpl}$ and by the small ratio $\lambda_\phi/\xif^2 \sim 10^{-10}$ and can be neglected. On the other hand, the contribution from Higgs-inflaton  coupling alone is of the same form as the original inflationary potential  and so can be absorbed in the redefinition of $\lambda_\phi$. This leaves the cross term between the two

\begin{equation}
  V_\mathrm{CW}\simeq \frac{\lfh m_\Omega^2}{32\pi^2}\left(\frac{\phi_J}{\Omega}\right)^2\log\left(\frac{m_h^2}{\mu^2}\right).
\end{equation}
Requiring that this contribution is subdominant gives the constraint

\begin{equation}
  \frac{\lfh}{\xif} \ll 72\pi^2e^{2\sqrt{\frac{2}{3}}\frac{\phi}{\Mpl}}.
\end{equation}
Thus, radiative corrections to the inflationary potential are always negligible, being suppressed by the exponential flatness of the potential.

\subsection{Vacuum stability}

In the Standard Model the self-coupling of the Higgs $\lambda_h$ turns negative above the critical scale $\mu_c\sim 10^{10}$ GeV. In order for the vacuum to remain stable against inflationary amplification of Higgs fluctuations the Higgs field must be heavy.  The effective potential for the Higgs is\footnote{Note that it is scaled by $\Omega^{-2}$ rather than $\Omega^{-4}$ because of non-minimal kinetic coupling. Alternatively one can think of $\Omega^{-4}V_h$ as the effective potential for the canonical variable $\tilde h$.}

\begin{eqnarray}
  V_\mathrm{eff} \simeq \Omega^{-2} V_h =\frac{1}{2}\lambda_{h\phi}\frac{\phi_J^2}{\Omega^2}h^2 + \frac{\lambda_h}{4\Omega^2}h^4.
\end{eqnarray}
The coupling to the inflaton introduces a large effective mass for the Higgs and the new instability scale defined as the location of the potential barrier becomes $h_c^2 = |\lambda_h|^{-1}\lambda_{h\phi}\phi_J^2$. The condition for $V''=0$ of the vanishing fluctuation mass is $h_c^2/3$. The Higgs is heavy during inflation if

\begin{equation}
  \frac{m_h^2}{H^2} \simeq 12\frac{\lambda_{h\phi}}{\lambda_\phi}\left(\frac{\phi_J}{\Omega}\right)^{-2}\Mpl^2 \gg 1.
\end{equation}
As long as this is satisfied and initially $3h_0^2 < h_c^2$ we have sufficient conditions for stability. Requiring that the Higgs is heavy gives the constraint

\begin{equation}
  \frac{\lfh}{\xi} \gg  \frac{6\pi^2\mathcal P_\zeta}{N^2} \approx 4\times 10^{-11}
\end{equation}
where the estimate is for around $55$ e-folds. Thus, we find that the coupling to the inflaton successfully stabilizes the vacuum without introducing radiative corrections to the inflationary potential.

\section{\label{sec:after}Higgs after inflation}

After inflation is over the inflaton starts to oscillate, first with a quadratic and then with a quartic potential of Equation~\eqref{eq:potential}. The equation of motion for the Higgs modes in the Hartree approximation is

\begin{equation}
 \label{eq:Higgseom}
   \ddot h_\mathbf{k} +\left(3H - \partial_t\log\Omega^2\right)\dot h_\mathbf{k}+\left[ \frac{k^2}{a^2} + \lfh \left(\frac{\phi_J}{\Omega}\right)^2 +3\lambda_h(\bar h)\left(\frac{\bar h}{\Omega}\right)^2\right] h_\mathbf{k} = 0
\end{equation}
where $\bar h \equiv \sqrt{\langle h^2\rangle}$. The coupling to the inflaton induces an oscillating mass term for the Higgs which results in amplification of Higgs fluctuations via parametric resonance~\cite{Kofman:1994rk,Kofman:1997yn}.

\subsection{Constraints from the quartic stage}
\label{sec:quartic}

Before we consider the oscillation in the quadratic regime let us examine what happens once the inflaton oscillation reaches the quartic regime. This happens when $\phi \sim \Mpl/\xi$. At this point the Einstein and Jordan frames agree and the dynamics can be described by the usual $\phi^4$ theory preheating~\cite{Greene:1997fu}. The expansion of the universe is radiation-domination-like and the situation is conformal to the Minkowski case. Changing to conformal time and rescaling both fields by $a$, the inflaton oscillates as

\begin{equation}
 \tilde \phi \sim \frac{\Mpl}{\xi}\operatorname{cn}\left(\sqrt{\lambda_\phi}\frac{\Mpl}{\xi}\tau; \frac{1}{\sqrt{2}}\right),
\end{equation}
where $\operatorname{cn}(x;k)$ is the Jacobi elliptic cosine, while the Higgs modes obey the equation

\begin{equation}
 \tilde h_k'' + \left[k^2 + \lfh \tilde \phi^2 +3\lambda_h(\bar h)\langle \tilde h^2\rangle\right]\tilde h_k \simeq 0.
\end{equation}
The resonance parameter is now

\begin{equation}
  q \sim \frac{\omega_h^2}{\omega_\phi^2} \sim \frac{\lfh}{\lambda_\phi} \simeq \frac{10^{10}}{5}\times \frac{\lfh}{\xi^2}.
\end{equation}
Since the effects of expansion factor out, modes don't exit resonance bands and so the production cannot be shut down by the expansion of the universe. Therefore if resonance is significant nothing can stop the Higgs from being destabilized unless inflaton decays before this via other interactions. Therefore we get a constraint

\begin{equation}
  \frac{\lfh}{\xi} < 5\times 10^{-10} \xi.
\end{equation}
Recall that stability during inflation required that $\lfh/\xi \gg 4\times  10^{-11}$. Thus, for large $\xi \gg 1$ this still leaves a sizeable window for stability both during inflation and the quartic stage. Now let us consider the intermediate, quadratic regime.

\subsection{Constraints from the quadratic stage}
\label{sec:quadratic}

\begin{figure}
  \includegraphics[width=\textwidth]{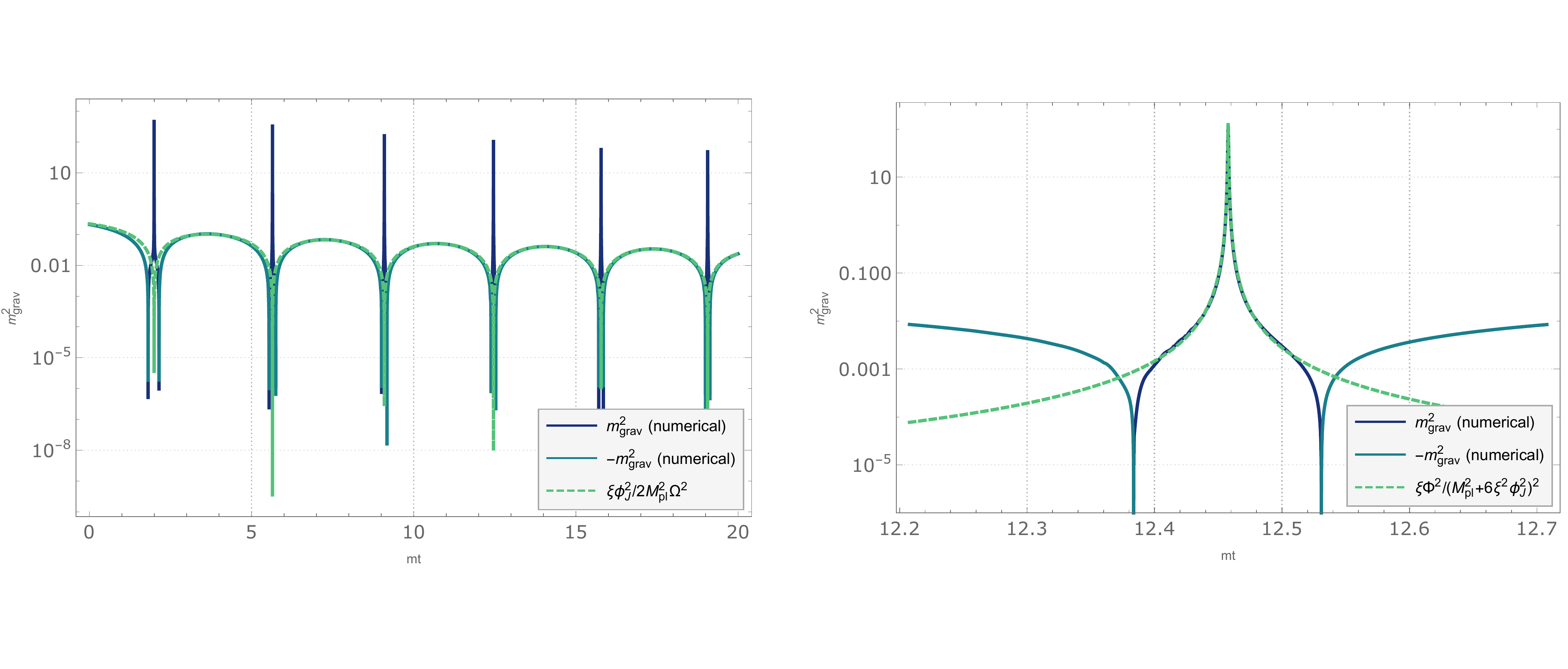}
  \caption{\label{fig:mgrav2}The behaviour of the gravitationally induced effective mass $m_\mathrm{grav}^2$. Far away from the zero-crossing the contribution is tachyonic while in its immediate vicinity the appears a large spike-like feature. Solid lines indicate the full numerical solution while the dashed lines correspond to the analytic estimates of equation~\eqref{eq:mgrav2}. The right panel shows a close vicinity around the fourth zero-crossing (when corrections due to the conformal factor have decayed sufficiently).}
\end{figure}

In the quadratic regime the Einstein frame inflaton oscillates in a harmonic potential with effective mass $m^2\equiv \frac{\lambda_\phi\Mpl^4}{3\xif^2}$. We define a new time variable $\tau \equiv mt$. The Jordan frame inflaton, which is what the Higgs is coupled to, then can be approximated by

\begin{equation}
   \label{eq:phifar}
  \phi_J^2 \simeq \sqrt{\frac{2}{3}}\frac{\Phi\Mpl}{\xif}|\sin\tau| ,\qquad \text{with} \qquad \Phi = \frac{2\sqrt{2}\Mpl}{\sqrt{\pi}\left(\frac{3}{2}+\tau\right)}
\end{equation}
being the amplitude of the Einstein frame inflaton field. The equation of motion for the rescaled Higgs fluctuations $X_k\equiv m^{1/2}\Omega^{-1} a^{3/2} h_k$ is

\begin{equation}
 X_\mathbf{k}''  + \omega_k^2X_\mathbf{k} = 0, \qquad \omega_k^2 = \frac{k^2}{m^2a^2} + m_{\phi h}^2 + m_\mathrm{grav}^2
\end{equation}
where the effective masses induced respectively by Higgs-inflaton interaction and gravity are

\begin{equation}
  m_{\phi h}^2 = \frac{\lfh}{\xif} \frac{\xif}{m^2}\left(\frac{\phi_J}{\Omega}\right)^2, \qquad m_\mathrm{grav}^2 = \partial_\tau^2\log\frac{\Omega}{a^{3/2}} - \left(\partial_\tau\log\frac{\Omega}{a^{3/2}}\right)^2.
\end{equation}
Note that these are measured in units of $m$. We have separated the ratio $\lfh/\xif$ because the quantity $\xif \phi_J^2$ behaves independently of $\xif$ for large non-minimal coupling and so this ratio will determine the strength of particle production.  The gravitationally induced mass $m_\mathrm{grav}$ can be estimated to be (see Appendix~\ref{sec:mgrav} for details)

\begin{equation}
  \label{eq:mgrav2}
  m_\mathrm{grav}^2\simeq \left\{\begin{array}{rcl}
  	-\frac{1}{2}\xif\Mpl^{-2}\Omega^{-2}\phi_J^2, &\qquad& 6\xif^2\phi_J^2 \gg \Mpl^2
  	\\
  	\frac{\xif\Phi^2\Mpl^2}{(\Mpl^2+6\xif^2\phi_J^2)^2}, &\qquad& 6\xif^2\phi_J^2 \ll \Mpl^2
  \end{array}\right. .
\end{equation}
Far away from the zero-crossing it contributes a tachyonic term. However, for the parameter range relevant for stability, $\lfh/\xif \gtrsim 10^{-10}$, the mass induced by the Higgs-inflaton coupling dominates and so $m_\mathrm{grav}$ can be neglected. In contrast, close to the zero-crossing, when $m_{\phi h}$ is small, the gravitationally induced term exhibits a large spike-like feature $\sim \xif \Phi^2/\Mpl^2$. The behaviour is illustrated in Figure~\ref{fig:mgrav2}. This can lead to particle production even in the absence of an explicit coupling to the inflaton as was pointed out in~\cite{Ema:2016dny}. 
As we shall see, however, destabilization occurs around $\lfh/\xif > 10^{-8}$. For such values effect of the spike is too brief and is well within the region adiabaticity violation so that it does not change the evolution of the modes. We have checked numerically that $m_\mathrm{grav}$ does not alter the dynamics for the coupling ranges relevant for destabilization. Thus, in the present case, this effect is unimportant and particle production is dominated by parametric resonance due to the Higgs-inflaton coupling.

\subsubsection*{Parametric resonance}

\begin{figure}
 \centering
  \includegraphics[width=1\textwidth]{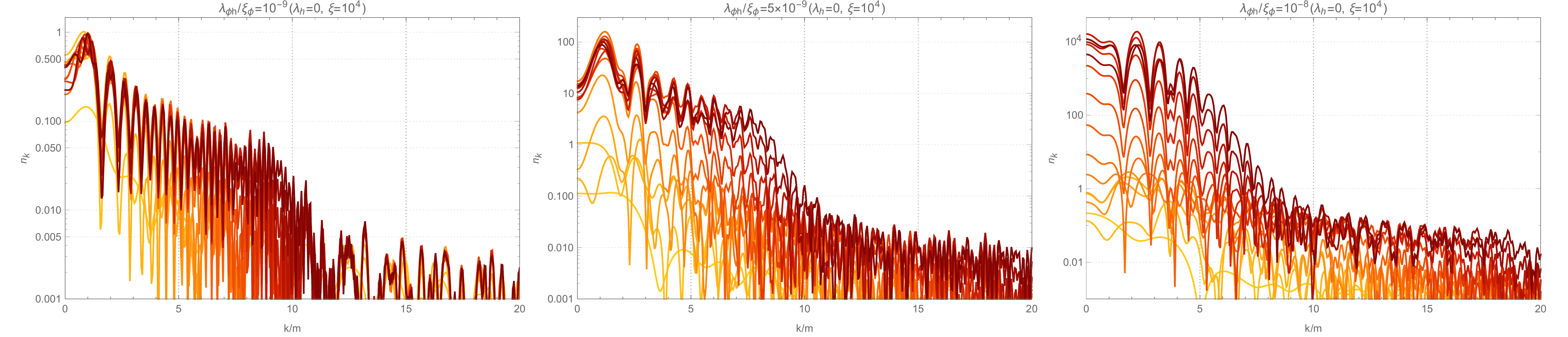}
  \caption{\label{fig:spectra_no_lh}Spectra at the first $30$ zero-crossings for different resonance parameters $\lfh/\xi$ with $\lambda_h=0$. Yellow indicates early times, red late times.}
\end{figure}

\begin{figure}
 \centering
  \includegraphics[width=.47\textwidth]{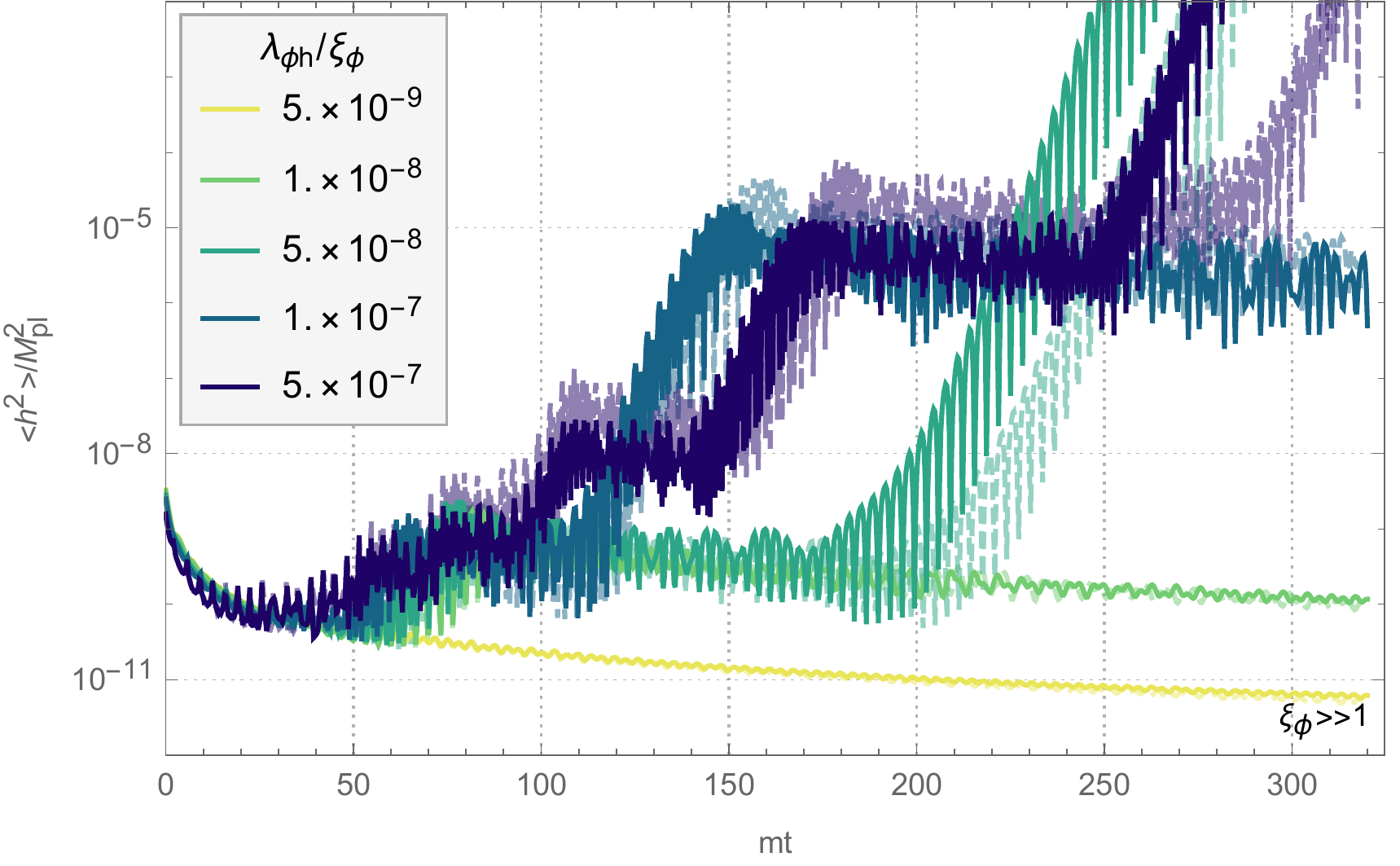}
   \includegraphics[width=.47\textwidth]{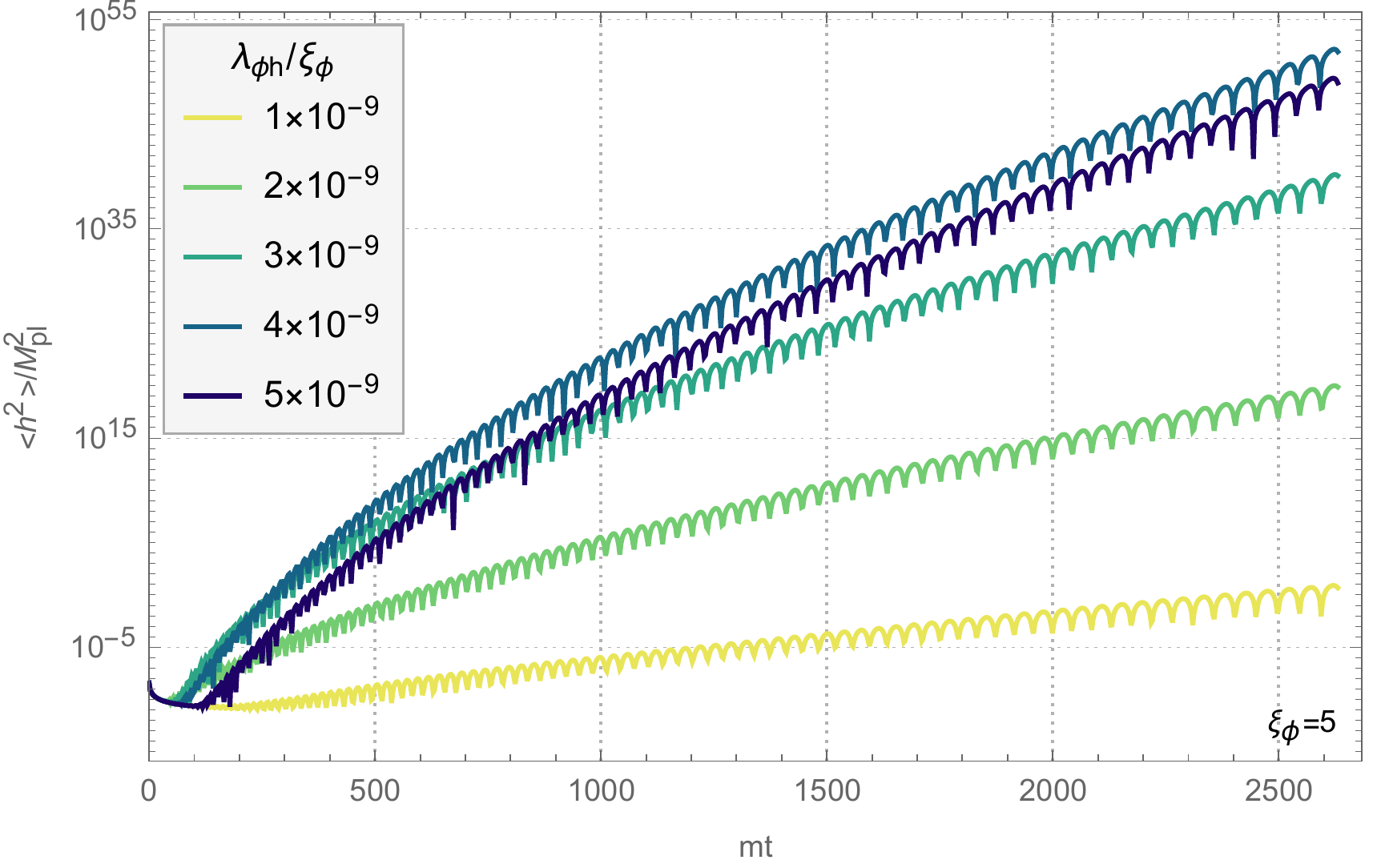}
  \caption{\label{fig:h2_no_lh}$\langle h^2\rangle$ for different resonance parameters $\lfh/\xi$ with $\lambda_h=0$. In the left panel, solid lines denote $\xif =10^4$ and dashed lines $\xif=10^6$, showing that for large non-minimal coupling preheating dynamics is indeed largely insensitive to $\xif$ and is determined by the ratio $\lfh/\xif$. The right panel shows the case of small non-minimal coupling $\xif=5$ where the system enters the quadratic regime early and Higgs modes continue to be amplified as discussed in Section~\ref{sec:quartic}, even for rather modest values $\lfh$.}
\end{figure}

The production of Higgs quanta proceeds via the non-perturbative process of broad parametric resonance which has been extensively studied in the literature~\cite{Kofman:1994rk,Kofman:1997yn,Greene:1997fu}. The case of non-minimal coupling to gravity has been considered in~\cite{Bezrukov:2008ut,GarciaBellido:2008ab} in the context of Higgs inflation and decay into gauge bosons and in~\cite{DeCross:2015uza,DeCross:2016fdz,DeCross:2016cbs} for general multi-field models. Far away from the zero-crossings of the inflaton the Higgs modes can be approximated by a WKB solution, so that before the $j$th zero-crossing

\begin{equation}
  \label{eq:before_j}
  X_\mathbf{k}(\tau) \simeq  \frac{\alpha_\mathbf{k}^j}{\sqrt{2\omega_k}}e^{-i\int^{\tau}\ud t\:\omega_k} + \frac{\beta_\mathbf{k}^j}{\sqrt{2\omega_k}}e^{i\int^{\tau}\ud \tau\:\omega_k}
\end{equation}
and after the zero crossing an equivalent one with $\alpha_k^{j+1}, \beta_k^{j+1}$. The mapping between the Bogolyubov coefficients can be obtained by solving the equations of motion near the zero-crossing and then matching the solution to the incoming and outgoing WKB modes. See Appendix~\ref{sec:WKBmapping} for more details. The occupation numbers after the $j$th zero crossing can be shown to be

\begin{equation}
  n_k^{j+1} = C_k^j + \left(1+2C_k^j\right)n_k^j -2\sin\theta_\mathrm{tot}\sqrt{C_k^j\left(1+C_k^j\right)}\sqrt{n_k^j\left(1+n_k^j\right)}
\end{equation}
where

\begin{equation}
 C_k^j \equiv \pi^2(A'A+B'B)^2,\qquad \theta_\mathrm{tot} \equiv 2\theta_k^j + \frac{4\kappa_j^3}{3q_j} +  \operatorname{arctan}\left(\frac{A'B+AB'}{AA'-BB'}\right) + \operatorname{arg}\beta_k^j  -  \operatorname{arg}\alpha_k^j.
\end{equation}
Here $A$ and $B$ refer to the Airy functions

\begin{equation}
  A\equiv \operatorname{Ai}\left(-\frac{\kappa_j^2}{q_j^{2/3}}\right), \quad B\equiv \operatorname{Bi}\left(-\frac{\kappa_j^2}{q_j^{2/3}}\right), \quad q_j \equiv \frac{4}{\sqrt{3\pi}}\frac{\Mpl^2}{m^2}\left(\frac{3}{2}+\tau_j\right)^{-1}\frac{\lfh}{\xif}, \quad \kappa_j^2 \equiv \frac{k^2}{m^2a_j^2},
\end{equation}
prime denotes a derivative, and $\theta_k^j \equiv \int^{\tau_j}\ud\tau\omega_k$ is the phase accrued by the $j$th zero-crossing. The numerically obtained occupation number spectra are plotted in Figure~\ref{fig:spectra_no_lh}.
In the large occupation number limit we can estimate $n_k^{j+1} \simeq \frac{1}{2}e^{2\pi\sum_{i=1}^j\mu_k^i}$ where the Floquet index is given by

\begin{equation}
  \mu_k^j \equiv \frac{1}{2\pi}\log\left(\frac{n_k^{j+1}}{n_k^j}\right) \simeq  \frac{1}{2\pi}\log\left[1+2C_k^j - 2\sin\theta_\mathrm{tot}\sqrt{C_k^j\left(1+C_k^j\right)}\right].
\end{equation}
The phase term $\theta_\mathrm{tot}$ can result in either enhancement of particle production or even in a decrease of particle numbers. However, it effectively behaves stochastically for large $\lfh/\xif$ in analogy to the $m^2\phi^2$ case~\cite{Kofman:1997yn}. Therefore, we take it to be zero on average so that

\begin{equation}
  n_k^{j+1} \simeq \frac{1}{2}e^{\sum_i^j\log\left(1+2C_k^i\right)}.
\end{equation}
With the use of the saddle-point approximation we can then estimate the variance of Higgs fluctuations (see Appendix~\ref{sec:Higgsvariance} for details)

\begin{equation}
  \langle h^2\rangle \simeq \frac{m^2\Omega^2}{2\pi^2a^3}\int\ud\kappa\:\kappa^2\frac{n_k}{\omega_k} \simeq 2\times 10^{-3}\times \sqrt{q}m^2\left(\frac{a}{a_0}\right)^{-3}\left(\frac{5}{3}\right)^j
\end{equation}
where $q=q_jj$ is the initial value of the resonance parameter. The numerical solution for the Higgs variance in the absence of self-interaction can be seen in Figure~\ref{fig:h2_no_lh}.

\subsection{Stability}

\begin{figure}
 \includegraphics[width=\textwidth]{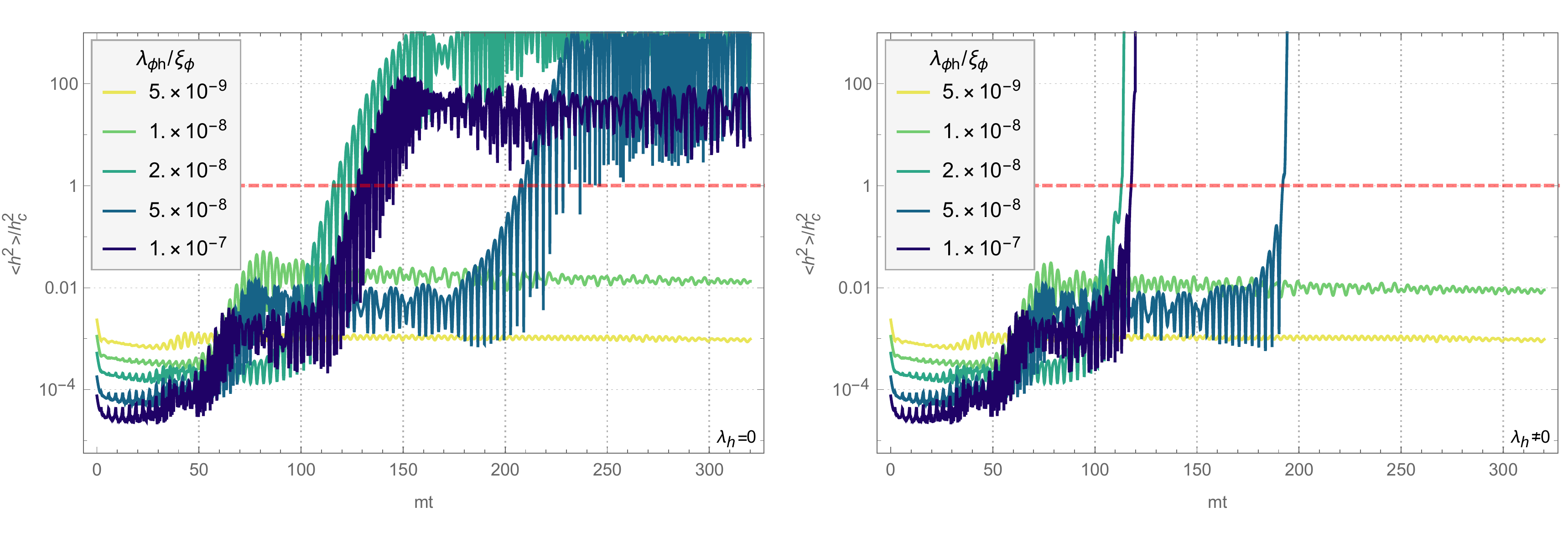}
 \caption{\label{fig:h2overhc2}Comparison to the Higgs variance to the location of the potential barrier. The left panel shows only the amplification from parametric resonance. The right panel shows the case where the Higgs self-coupling was taken into account.}
\end{figure}

The Higgs fluctuations are amplified exponentially due to parametric resonance induced by the Higgs-inflaton coupling. If this continues long enough the Higgs will be driven into the true vacuum at large field values. Note that even for $q\sim 1$, $\langle h^2\rangle^{1/2}$ is above the instability scale of the pure standard model $\mu_c^\mathrm{SM}\sim 10^{-10}$ GeV so that the contribution of the Higgs self-coupling is negative ($\lambda_h<0$) throughout preheating, contributing a tachyonic mass to the Higgs modes. The vacuum remains stable as long as the resonance terminates before this contribution overpowers the stabilizing effect of the Higgs-inflaton coupling. Taking into account the self-coupling at the level of Hartree approximation, the frequency of Higgs modes around the zero-crossing is

\begin{equation}
 \omega_k^2 = \frac{\kappa^2}{a^2}+\Omega^{-2}m^{-2}\left(\lambda_{h\phi}\phi_J^2 - 3|\lambda_h|\langle h^2\rangle \right) \simeq \frac{\kappa_j^2}{a_j^2} + q_j |\Delta\tau| - 3|\lambda_h|\frac{\langle h^2\rangle}{m^2}.
\end{equation}
Near the zero-crossing the stabilizing effect of the inflaton temporarily disappears and the tachyonic term due to self-interaction dominates for the time interval

\begin{equation}
  \Delta \tau_\mathrm{tach} = \frac{3|\lambda_h|\langle h^2\rangle}{q_jm^2} \equiv \frac{m_\mathrm{tach}^2}{q_jm^2}.
\end{equation}
For stability we require that $m_\mathrm{tach}\Delta \tau_\mathrm{tach}<1$ which translates to

\begin{equation}
 \label{eq:destabilization}
  \frac{m_\mathrm{tach}^3}{q_jm^3} <1 \qquad \mathrm{or} \qquad \langle h^2 \rangle < \frac{m^2q_j^{2/3}}{3|\lambda_h|}.
\end{equation}
Therefore the stability condition is

\begin{equation}
   j < j_\mathrm{inst} \equiv  \frac{1}{\log\left(\frac{5}{3}\right)}\log\left(\frac{q^{1/6}}{3|\lambda_h|A}\right).
\end{equation}
Requiring that the resonance terminates before the instability condition becomes violated $j_\mathrm{end} \sim q < j_\mathrm{inst}$ we get

\begin{equation}
   q < \frac{1}{\log\left(\frac{5}{3}\right)}\log\left(\frac{q^{1/6}}{3A|\lambda_h|}\right) \qquad \Rightarrow \qquad q <20
\end{equation}
This translates to a constraint on couplings

\begin{equation}
 \frac{\lambda_{h\phi}}{\xif} < \frac{\sqrt{3\pi^3}}{4}\frac{m^2}{\Mpl^2}q_\mathrm{crit} \approx 8\times 10^{-9}
\end{equation}
where we have taken $|\lambda_h|\sim 10^{-2}$. Meanwhile, recall that the stability condition during inflation was

\begin{equation}
  \frac{\lfh}{\xif} \gg 4\times 10^{-11}.
\end{equation}
Thus, only the range $\lfh/\xif\sim 10^{-10}..10^{-8}$ is viable in terms of stability both during and after inflation. A numerical comparison of $\langle h^2\rangle$ to the instability scale can be seen in Figure~\ref{fig:h2overhc2}. We can observe that self-coupling leads to explosive destabilization; however, in terms of whether $\langle h^2\rangle$ exceeds the barrier it does result in a qualitative change.

\begin{figure}[t]
 \includegraphics[width=\textwidth]{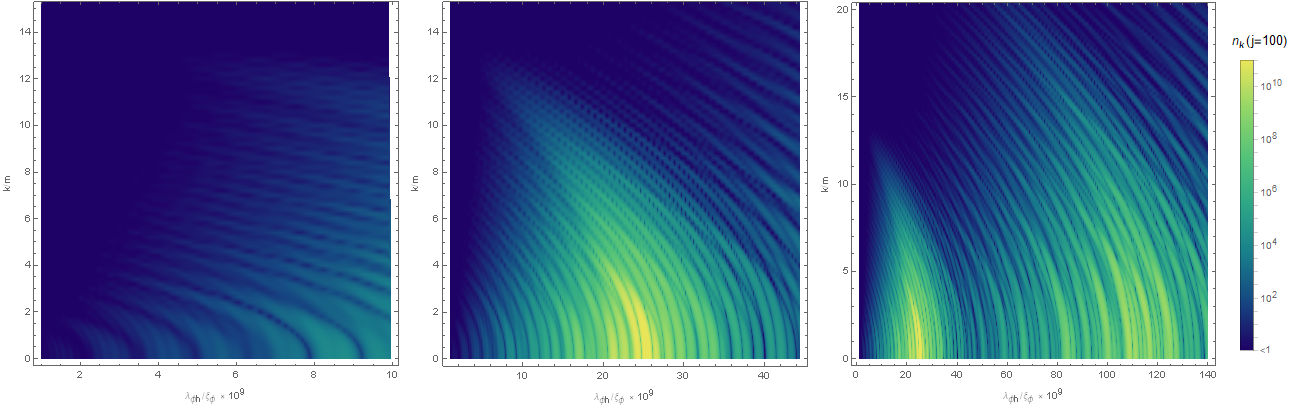}
 \caption{\label{fig:stabilitychart}Occupation numbers after $100$ zero-crossings. The interplay of phases causes constructive and destructive interference making the strength of resonance non-monotonic. For example, particle production is very strong around $\lfh/\xif \sim 2\times 10^{-8}$ and less so for higher coupling values. However, for higher coupling values the resonance also lasts much longer (beyond $100$ zero-crossings).}
\end{figure}

In our analytic estimates, we have not taken into account the phase factor which makes the dependence of the resonance on coupling non-monotonic. Numerical solution shows regions of enhanced  and diminished resonance in the parameter space, as can be seen in Figure~\ref{fig:stabilitychart}. In particular, resonance is very strong for $\lfh/\xif\sim 2\times 10^{-8}$ and weaker for larger couplings. However, resonance continues as long as $q_j>1$ which decays rather slowly, $q_j\propto j^{-1}$, as a result of the fact that the Higgs is coupled to $\phi_J$ rather than the $\phi$ field\footnote{This is in contrast to the minimally coupled $m^2\phi^2$ case where $q_j\propto j^{-2}$.}. Therefore the duration of resonance is proportional to $\lfh/\xif$ and so destabilization occurs even in the weaker regions of Figure~\ref{fig:stabilitychart} as long as $\lfh/\xif>10^{-8}$. We have also checked that the effect of $m_\mathrm{grav}$ is unimportant.

\subsection{Backreaction}

So far we have considered the case where the resonance is shut down by the expansion of the universe. If enough Higgs quanta are produced, however, they can backreact on the inflaton dynamics, changing its equation of motion. For large couplings this can shut down the resonance while it is still broad. In this section we will estimate the significance of backreaction for the results presented above. The gradient of the inflaton potential is 

\begin{equation}
\label{eq:backreaction}
  V'(\phi) \simeq \frac{\lambda_\phi\phi_J^3 + \xif \phi_J \sigma_\phi^2}{\sqrt{1+6\xif^2\phi_J^2/\Mpl^2}}
\end{equation}
where the backreaction of Higgs fluctuations is given by

\begin{equation}
 \sigma_\phi^2 \equiv \frac{\lfh}{\xif}\langle h^2\rangle - \Mpl^{-2}\langle \partial_\mu h\partial^\mu h\rangle - 3\lambda_h\langle h^2\rangle^2.
\end{equation}
The derivative term $\langle \partial_\mu h \partial^\mu h\rangle \sim -\lfh \phi_J^2 \langle h^2\rangle$ is subdominant since during preheating we have $\xif\phi_J^2 \ll \Mpl^2$. Also the last term is subdominant until destabilization. From equation~\eqref{eq:destabilization}, at the time of destabilization $\langle h^2\rangle \sim q_j^{2/3}m^2/3|\lambda_h|$. The backreaction term in equation~\eqref{eq:backreaction} remains subdominant until destabilization as long as

\begin{equation}
  \frac{\lfh}{\xif} \lesssim 0.2\times \left(\frac{m}{\Mpl}\right)^{4/5}\tau_\mathrm{destab.}^{-1/5} \sim 10^{-5}.
\end{equation}
Therefore, backreaction is unimportant for the ranges of couplings discussed in the previous sections. However, one might consider the possibility that for very large couplings $\lfh \gg 10^{-5}\xif$ backreaction shuts down the resonance before destabilization can occur thus rescuing the Higgs. However, at the time of backreaction a significant part the inflaton condensate decays thus removing the stabilizing effect of the Higgs-inflaton coupling. Detailed investigation of such dynamics requires non-linear analysis using lattice simulations and is beyond the scope of the present work. Furthermore, as discussed in the next section, the large coupling regime raises questions of unitarity because large momenta are amplified by parametric resonance so that the model assumed above may not be sufficient to described the dynamics.

\subsection{Unitarity}

The subject of unitarity in non-minimaly coupled theories has gained much attention, especially in the context of Higgs inflation~\cite{Burgess:2009ea,Barbon:2009ya,Hertzberg:2010dc,Bezrukov:2010jz,Bezrukov:2012hx,Bezrukov:2013fka}. The UV cutoff of the theory depends on the amplitude of the background field so that when $\phi_J^2\gg \Mpl^2/\xif$ the cutoff is given by $\Lambda \sim \Mpl$ allowing for a robust effective field theory description~\cite{Bezrukov:2010jz,Bezrukov:2012hx,Bezrukov:2013fka}. In the small field regime $\phi_J\ll \Mpl/\xif$ the cutoff is $\Lambda \sim \Mpl/\xif$. In the intermediate regime the cutoff depends on the field amplitude, interpolating between these two values.

\begin{figure}[t]
\centering
  \includegraphics[width=.550\textwidth]{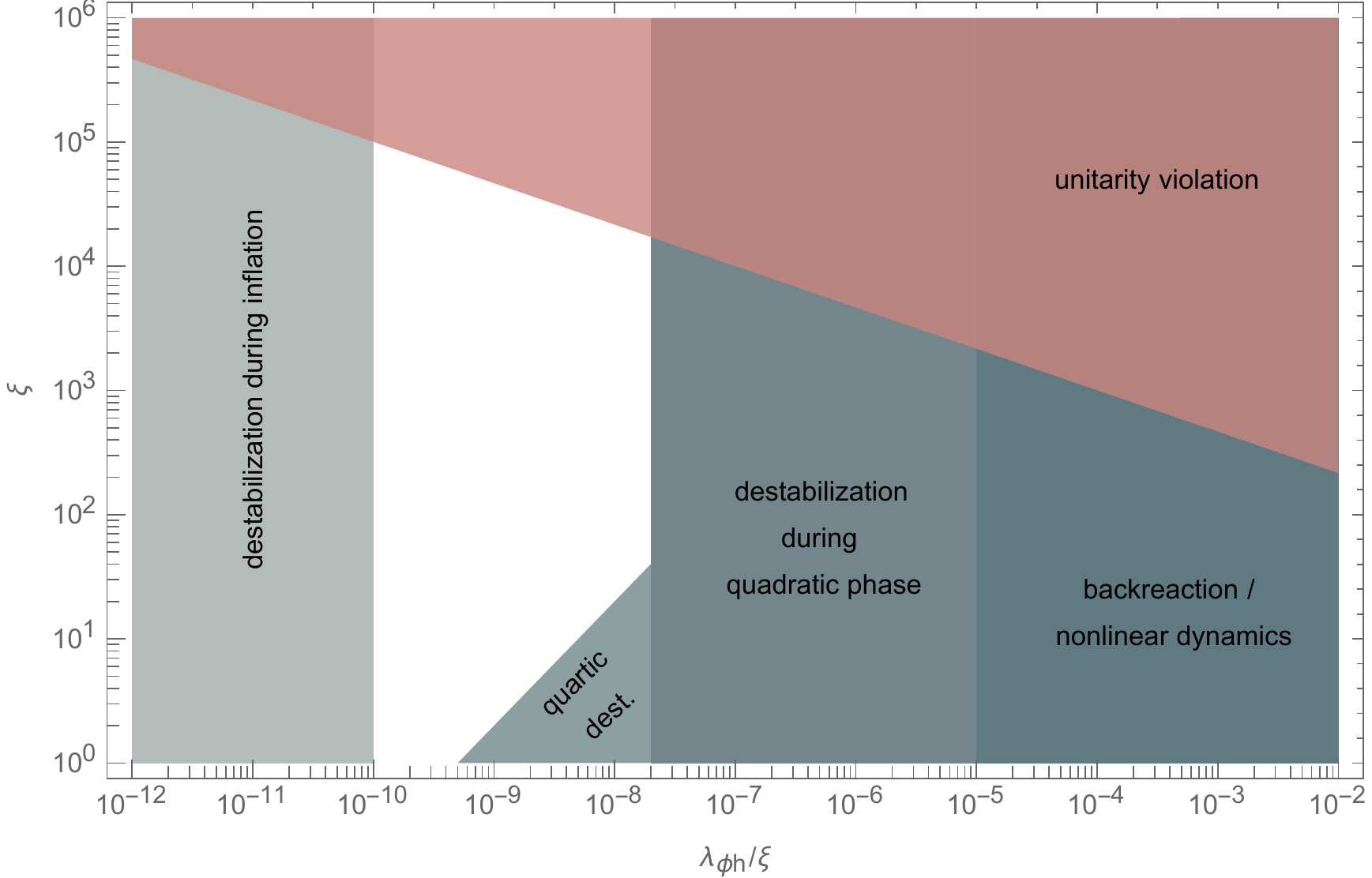}
  \caption{\label{fig:constraints}Constraints from stability during and after inflation. }
\end{figure}

As discussed in~\cite{Ema:2016dny,DeCross:2016fdz}, we take the effective theory description to be valid as long as momenta do not exceed the UV cutoff. Parametric resonance produces quanta with typical momenta $k\sim q^{1/3}m$. Requiring that these stay below the cutoff $\Mpl/\xif$ gives the following unitarity bound for the couplings:

\begin{equation}
\frac{\lfh}{\xif}< \xif^{-3}\left(\frac{\Mpl}{m}\right) \sim 10^{5} \times \xif^{-3}.
\end{equation}
Thus, unless $\xif$ is extremely large, unitarity is not violated for the ranges relevant for Higgs destabilization ($\lfh\sim 10^{-8}\xif$). Furthermore, this constraint may be too conservative because the Higgs stays subdominant and the amplitude of the inflaton is large $\phi_J \gg \Mpl/\xif$. Consequently the cutoff is also higher than $\Mpl/\xif$. On the other hand,  in the strong coupling regime where backreaction becomes important before destabilization the inflaton is expected to transfer most of its energy to Higgs quanta decreasing the amplitude of the field and lowering the cutoff. Therefore in this range a UV complete description may be necessary for obtaining the correct dynamics. The unitarity bound is shown in Figure~\ref{fig:constraints} where we summarize the model constraints. It should also be noted that in the region of large ratio $\lfh/\xi$ the portal coupling $\lfh$ becomes large, changing the running of $\lambda_h$, making the vacuum more stable. If it is increased further it becomes non-perturbative at low-energy which is problematic in terms of experimantally verified Higgs phenomenology.

\section{\label{sec:numerical}Numerical analysis}

To complement the analytic investigation of the previous sections, we have also studied the dynamics of the Higgs numerically. The equations of motion for the inflaton in the Einstein frame are

\begin{equation}
 \ddot \phi + 3H\dot\phi + \frac{\ud V_E}{\ud \phi} = 0, \qquad 3\Mpl^2 H^2 = \frac{1}{2}\dot\phi^2 + V_E , \qquad \dot H = -\frac{\dot\phi^2}{2\Mpl^2}
\end{equation}
However, because there is no closed form expression for the potential in terms of the Einstein frame field we instead solve the dynamics of the variable $\phi_J$, obeying the equations

\begin{equation}
\label{eq:phiJeom}
 \ddot\phi_J + \left(3H - \frac{1}{2}\frac{D'}{D}\dot\phi_J\right) \dot\phi_J + D \frac{\ud V_E}{\ud \phi_J} = 0, \qquad 3\Mpl^2H^2 = \frac{1}{2}D^{-1} \dot \phi_J^2 + V_E(\phi_J)
\end{equation}
where 

\begin{equation}
  D(\phi_J) \equiv \left(\frac{\ud\phi_J}{\ud\phi}\right)^2 =  \frac{\Omega^4}{\Omega^2 + 6\xif^2\phi_J^2/\Mpl^2}.
\end{equation}
Note, however, that time and the Hubble parameter are still measured in the Einstein frame. We solve the inflaton dynamics with $55$ e-folds of inflation and then use the obtained solution to get the evolution of the Higgs fluctuations whose equation of motion is given by~\eqref{eq:Higgseom}. We take the self-coupling of the Higgs into account at the Hartree level. This is implemented as follows. We solve the evolution of Higgs modes one inflaton half-cycle at a time for different momenta and then integrate to obtain $\langle h^2\rangle$. This is then used to evolve the modes across the next half-cycle of inflaton oscillation. This procedure is sufficient for finding the bounds from stability as  destabilization occurs long before backreaction and rescattering become important as long as $\lfh\xif<10^{-3}$.

\section{Conclusions}
\label{sec:conclusions}

We have studied the behaviour of the Standard Model Higgs in a theory of inflation where the inflaton is non-minimally coupled to gravity. This is a well-motivated model both theoretically, as non-minimal couplings are generically radiatively generated, and experimentally, with its cosmological prediction falling squarely in the middle of current observational bounds.

We find that the stability of the electroweak vacuum during and immediately after inflation severely constrains the possible parameter space. During inflation, the pure SM Higgs is a light  field whose fluctuations are amplified by \emph{de Sitter} expansion leading to destabilization. If, on the other hand, the Higgs is stabilized during inflation via coupling to the inflaton its fluctuations are amplified after inflation by parametric resonance resulting from an oscillating inflaton background. The dynamics of the inflaton is largely independent of the non-minimal coupling to gravity during inflation and a period of harmonic oscillation following it as long as the amplitude is rescaled by $\sqrt{\xi}$. As a result, the relevant coupling for the Higgs is the ratio $\lfh/\xif$. We find that early universe vacuum stability is realized if this ratio is in the range $\lfh/\xif\sim 10^{-10}..10^{-8}$.

During inflation a ratio of $\lfh/\xif \gtrsim 10^{-10}$ is needed in order to make the Higgs heavy and roll to the vacuum at low field values. During preheating with large $\xif$ the inflaton oscillates in harmonic potential and parametric amplification of Higgs quanta ends due to expansion before destabilization for $\lfh/\xif \lesssim 10^{-8}$. For moderate values of the non-minimal coupling the system soon transitions from a harmonic potential to quartic where the effects of the expansion factor out and so cannot end the resonance. It will instead be terminated by the backreaction of produced quanta.

The backreaction effects are not important when the couplings are in the above range. For very large couplings backreaction may come into play at comparable scales to the destabilization, necessitating the use of non-linear solutions including rescattering effects. However, in such a regime quanta with momenta higher that the UV cutoff of the theory are easily amplified indicating that a UV complete treatment may be necessary. This is beyond the scope of the current analysis. The constraints are summarized in Figure~\ref{fig:constraints}.

Since an inflaton field must couple to the SM in order to facilitate reheating Higgs portal coupling is expected to be a generic feature of scalar field models of inflation. Thus, the tight constraints imposed by vacuum metastability can in fact prove an important discriminator between inflationary models.

\subsection*{Acknowledgements}

I am grateful to Oleg Lebedev, Yohei Ema, Mindaugas Kar\v ciauskas, and Marco Zatta for useful discussions.

\appendix

\section{Gravitationally induced mass}
\label{sec:mgrav}

In this section we estimate the contribution to the Higgs effective mass from the derivatives of the conformal factor. Far away from its zero-crossings the inflaton is well approximated by~\eqref{eq:phifar}. Then the dominant contribution to the Higgs effective mass is trivially

\begin{equation}
  m_\mathrm{grav}^2 \simeq  \frac{1}{2}\partial_\tau^2\log\Omega^2 \simeq - \frac{1}{2}\frac{\xif\phi_J^2}{\Mpl^2\Omega^2}.
\end{equation}
Near the zero-crossing we may take $V\simeq 0$, $\Omega \simeq 1$, $D\simeq (1+6\xif^2\phi_J^2/\Mpl^2)^{-1}$. Then equation of motion~\eqref{eq:phiJeom} reduces to

\begin{equation}
 \frac{\sqrt{D}}{a^3}\frac{\ud}{\ud \tau}\left(\frac{a^3}{\sqrt{D}}\phi_J'\right) \simeq 0
\end{equation}
whose solution is $\phi_J' \simeq C\sqrt{D}$. Integrating this and matching it to the sinusoidal solution~\eqref{eq:phifar} in the limit $\sqrt{6}\xif\phi_J\gg\Mpl$ we obtain the solution around the $j$th zero-crossing

\begin{equation}
\label{eq:phiJcanonical}
  \sqrt{6}\xif\frac{\phi_J}{\Mpl}\sqrt{1+6\xif^2\frac{\phi_J^2}{\Mpl^2}} + \mathrm{arsinh}\left(\sqrt{6}\xif\frac{\phi_J}{\Mpl}\right) =  (-1)^{j}\frac{\Phi}{\Mpl}(\tau-\tau_j)
\end{equation}
from which we can then find the derivatives of the inflaton field

\begin{equation}
 \phi_J' \simeq (-1)^j\frac{\Phi\Mpl}{\sqrt{\Mpl^2+6\xif^2\phi_J^2}}, \qquad \phi_J'' \simeq - \frac{6\xif^2\Phi^2\Mpl^2\phi_J}{(\Mpl^2+6\xif^2\phi_J^2)^2}.
\end{equation}
Thus the first derivative exhibits a peak of hight $\sim \Phi$ at the zero-crossing and the second derivative has two peaks of hight $\sim \xi\Phi^2$ on either side of the zero-crossing. The total contribution of the Higgs effective mass in the limit $6\xif^2\phi_J^2 \ll \Mpl^2$ is then

\begin{equation}
  m_\mathrm{grav}^2 \simeq \frac{\xif\Phi^2\Mpl^2}{(\Mpl^2+6\xif^2\phi_J^2)^2}.
\end{equation}

\section{Mapping between WKB modes}
\label{sec:WKBmapping}

This section covers the parametric resonance results of Section~\ref{sec:quadratic}. Near the zero-crossing we can approximate the accrued phase

\begin{equation}
  \theta(\tau) = \int_0^\tau \ud\tau \omega_k \simeq \theta_k^j + \int_{0}^{\Delta\tau}\ud(\Delta\tau)\omega_k(\tau_j+\Delta \tau) \simeq  \theta_k^j + \mathrm{sign}(\Delta\tau)\frac{2}{3q_j}\left[\left(\kappa_j^2+q_j|\Delta\tau|\right)^{3/2}-\kappa_j^3\right]
\end{equation}
where $\Delta\tau \equiv \tau-\tau_j$ and $\theta_k^j \equiv \int^{\tau_j}\ud\tau\omega_k(\tau)$. The  equation of motion close to the zero-crossing is

\begin{equation}
  X'' + (\kappa_j^2 + q_j|\Delta \tau|)X\simeq 0 
\end{equation}
whose solutions are Airy functions. Before the zero-crossing these are

\begin{equation}
  X = C_1^- \operatorname{Ai}\left(\frac{q_j\Delta\tau -\kappa_j^2}{q_j^{2/3}}\right) + C_2^- \operatorname{Bi}\left(\frac{q_j\Delta\tau -\kappa_j^2}{q_j^{2/3}}\right) 
\end{equation}
and after the zero-crossing

\begin{equation}
  X = C_1^+ \operatorname{Ai}\left(-\frac{q_j\Delta\tau + \kappa_j^2}{(-q_j)^{2/3}}\right) + C_2^+ \operatorname{Bi}\left(-\frac{q_j\Delta\tau +\kappa_j^2}{(-q_j)^{2/3}}\right) 
\end{equation}
Far away from the zero crossing the solution is given by the WKB form~\eqref{eq:before_j}. Using the large-argument expansions for Airy functions 

\begin{eqnarray}
  \operatorname{Ai}\left(-\frac{\kappa_j^2-q_j\Delta \tau}{q_j^{2/3}}\right) & \simeq & \frac{q_j^{1/6}}{\sqrt{\pi}(\kappa_j^2-q_j\Delta\tau)^{1/4}}\sin\left[\frac{2}{3q_j}\left(\kappa_j^2-q_j\Delta\tau\right)^{3/2} + \frac{\pi}{4}\right]
  \\
  \operatorname{Bi}\left(-\frac{\kappa_j^2-q_j\Delta \tau}{q_j^{2/3}}\right) & \simeq & \frac{q_j^{1/6}}{\sqrt{\pi}(\kappa_j^2-q_j\Delta\tau)^{1/4}}\cos\left[\frac{2}{3q_j}\left(\kappa_j^2-q_j\Delta\tau\right)^{3/2} + \frac{\pi}{4}\right]   
\end{eqnarray}
and
\begin{eqnarray}
 \operatorname{Ai}\left(-\frac{q_j\Delta\tau + \kappa_j^2}{(-q_j)^{2/3}}\right) & \simeq & \frac{e^{-\frac{i\pi}{12}}q_j^{1/6}}{\sqrt{4\pi}}\frac{e^{-\frac{2i}{3q_j}(\kappa_j^2 + q_j\Delta\tau)^{3/2}}}{(\kappa_j^2+q_j\Delta\tau)^{1/4}}
 \\
  \operatorname{Bi}\left(-\frac{q_j\Delta\tau + \kappa_j^2}{(-q_j)^{2/3}}\right) & \simeq &\frac{e^{-\frac{i\pi}{12}}q_j^{1/6}}{\sqrt{4\pi}}\frac{2e^{\frac{2i}{3q_j}(\kappa_j^2 + q_j\Delta\tau)^{3/2}} + ie^{-\frac{2i}{3q_j}(\kappa_j^2 + q_j\Delta\tau)^{3/2}}}{(\kappa_j^2+q_j\Delta\tau)^{1/4}}
\end{eqnarray}
and matching the incoming and outgoing WKB modes one can obtain the mapping between Bogolyubov coefficients before and after the zero-crossing

\begin{equation}
  \left(\begin{array}{c}
  	\alpha_k^{j+1}\\ \beta_k^{j+1}
  \end{array}\right)
  = \mathcal M \left(\begin{array}{c}
  	\alpha_k^{j}\\ \beta_k^{j}
  \end{array}\right)
\end{equation}
with the relevant components given by

\begin{eqnarray}
 \mathcal M_{\beta\alpha} &  = &   -i \pi e^{-2i\theta_k^j}(AA'+BB')
 \\
 \mathcal M_{\beta\beta} & = & - \pi e^{\frac{4i\kappa_j^3}{3q_j}}\Big[(AA'-BB') + i(A'B+AB')\Big]
\end{eqnarray}
where $A$ and $B$ referring to the two Airy functions with argument $-\kappa_j^2/q_j^{2/3}$. We now have 

\begin{equation}
 C_k^j \equiv \left|\mathcal M_{\beta\alpha}\right|^2 = \pi^2(AA'+BB')^2, \qquad \left|\mathcal M_{\beta\beta}\right|^2 = 1+\pi^2(AA'+BB')^2 = 1+C_k^j
\end{equation}
and

\begin{equation}
 \operatorname{arg}\mathcal M_{\beta\alpha} = -2\theta_k^j - \frac{\pi}{2}, \qquad \operatorname{arg}\mathcal M_{\beta\beta} = \frac{4\kappa_j^3}{2q_j} + \operatorname{arctan}\left(\frac{A'B+AB'}{AA'-BB'}\right)
\end{equation}
and the occupation numbers after the zero-crossing are

\begin{equation}
  \left|\beta_k^{j+1}\right|^2 \equiv  n_k^{j+1} = C_k^j + \left(1+2C_k^j\right)n_k^j -2\sin\theta_\mathrm{tot}\sqrt{C_k^j\left(1+C_k^j\right)}\sqrt{n_k^j\left(1+n_k^j\right)}
\end{equation}
where

\begin{equation}
  \theta_\mathrm{tot} \equiv 2\theta_k^j + \frac{4\kappa_j^3}{3q_j} +  \operatorname{arctan}\left(\frac{A'B+AB'}{AA'-BB'}\right) + \operatorname{arg}\beta_k^j  -  \operatorname{arg}\alpha_k^j.
\end{equation}

\section{Calculation of the Higgs variance}
\label{sec:Higgsvariance}

This section describes the details of Higgs production during parametric resonance. Taking the typical value for the Floquet index, corresponding to $\theta_\mathrm{tot}=0$, the occupation numbers after $j$ zero-crossings are

\begin{equation}
  n_k^{j+1} \simeq \frac{1}{2}e^{\sum_i^j\log\left(1+2C_k^i\right)}
\end{equation}
The window function $C_k^j$ has a maximum at $z=0$

\begin{equation}
  C_m \equiv C(z=0) = \left[\frac{2\pi}{3\Gamma(1/3)\Gamma(2/3)}\right]^2 =  \frac{1}{3}
\end{equation}
and the derivative is

\begin{equation}
  D_m \equiv 2 C'(z=0) = -\frac{16\pi}{3^{7/6}\left[\Gamma(1/3)\right]^2} \approx -1.94403
\end{equation}
We may expand the exponent of the occupation number around this maximum

\begin{eqnarray}
 \sum_i^{j}\log(1+2C_k^j) & \simeq & j\log\left(1+2C_m\right) + \frac{D_m}{1+2C_m}\sum_{i=1}^j\frac{\kappa_i^2}{q_i^{2/3}}
  \\
 &=& j\log\left(\frac{5}{3}\right) + \frac{3D_m \kappa^2}{5q^{2/3}a_0^2}\left(\sum_{i=1}^j\frac{1}{i^{2/3}}\right) \equiv j\log\left(\frac{5}{3}\right) + \frac{3D_m \kappa^2}{5q^{2/3}a_0^2}S(j)
\end{eqnarray}
where $q\equiv j q_j$ and $a_0 \equiv a_j j^{-2/3}$. The occupation number is

\begin{equation}
  n_k^{j+1} \simeq \frac{1}{2}\left(\frac{5}{3}\right)^j e^{-\bar\mu_j \kappa^2}, \qquad \bar\mu_j \equiv \frac{3|D_m| }{5a_0^2q^{2/3}}S(j) 
\end{equation}
and where $\kappa\equiv k/m$. Then the dispersion of Higgs fluctuations far away from the zero-crossing can be obtained as

\begin{equation}
 \langle h^2\rangle = \frac{\Omega^2}{a^3m}\int\frac{\ud^3k}{(2\pi)^3}|X_k|^2 \simeq \frac{\Omega^2m^2}{2\pi^2}\int \ud \kappa \kappa^2\frac{n_k}{\omega_k} \simeq \frac{m^2\Omega^2}{4\pi^2a^3}\left(\frac{5}{3}\right)^j\int\ud\kappa\frac{\kappa^2e^{-\bar\mu_j \kappa^2}}{\omega_k}
\end{equation}
We estimate the integral using the saddle-point method. Write $\kappa^2e^{-\bar\mu_j \kappa^2} = e^{f(\kappa)}$. Expanding around the maximum $\kappa_\mathrm{max}^2 = \bar\mu_j^{-1}$ we have

\begin{equation}
  f(\kappa) \simeq f(\kappa_\mathrm{max}) + \frac{1}{2}f''(\kappa_\mathrm{max})(\kappa-\kappa_\mathrm{max})^2 = \log\kappa_\mathrm{max}^2-1 - 2\bar\mu_j(\kappa-\kappa_\mathrm{max})^2 
\end{equation}

\begin{equation}
 \int \ud \kappa \frac{\kappa^2e^{-\bar\mu_j\kappa^2}}{\omega}\simeq \frac{\kappa_\mathrm{max}^2 e^{-1}}{\omega(\kappa_\mathrm{max})}\int_{-\infty}^\infty \ud\tilde \kappa e^{-2\bar\mu_j \tilde\kappa^2} = \sqrt{\frac{\pi}{2e^2}}\frac{\kappa_\mathrm{max}^3}{\omega(\kappa_\mathrm{max})}.
\end{equation}
The dispersion of fluctuations is then

\begin{equation}
  \langle h^2\rangle \simeq \frac{m^2\kappa_\mathrm{max}^3\Omega^2}{2^{5/2}e\pi^{3/2}a^3\omega_k(\kappa_\mathrm{max})}\left(\frac{5}{3}\right)^j
\end{equation}
with

\begin{equation}
 S(j) =  \sum_{i=1}^j i^{-2/3} \simeq \frac{3}{2^{1/3}}\left[\left(2j+1\right)^{1/3} - 1\right] \simeq 3\left(\frac{a_j}{a_0}\right)^{1/2}
\end{equation}
so that 

\begin{equation}
 \label{eq:mubar}
 \bar\mu_j \simeq \frac{3|D_m| }{5q^{2/3}a_0^2}S(j) \simeq \frac{9|D_m|}{5 q^{2/3}a_0^2} \left(\frac{a_j}{a_0}\right)^{1/2}
\end{equation}
The frequency of the modes is

\begin{equation}
\label{eq:freq}
  \omega_k^2 = \left[\frac{\kappa_j^2}{a_j^2}+ q_j|\sin(mt)|\right]
\end{equation}
The momentum cutoff can be identified from the window function as

\begin{equation}
  \kappa_\mathrm{cut}^2 \sim q_j^{2/3}a_j^2  \sim q^{2/3}a_0a_j
\end{equation}
Thus for broad resonance $q_j \gg 1$ the mass term dominates over the momentum contribution:

\begin{equation}
  \omega_k^2 \simeq q_j|\sin(mt)| \simeq q_j
\end{equation}
Therefore, the final estimate for the Higgs variance is

\begin{equation}
  \langle h^2\rangle \simeq A \sqrt{q}m^2\left(\frac{a}{a_0}\right)^{-3}\left(\frac{5}{3}\right)^j,\qquad A   \equiv \left(\frac{5}{9|D_m|}\right)^{3/2}\frac{1}{2^{5/2}e\pi^{3/2}} \approx 2\times 10^{-3}
\end{equation}

\bibliography{refs}
\bibliographystyle{JHEP}

\end{document}